\documentstyle[aps,prb,fancyheadings,epsf,multicol,amssymb]{revtex}
\newcommand{\lrule}{ \noindent
        \rule{0.5\textwidth}{0.1mm}\rule{0.1mm}{3pt}\newline }
\newcommand{\rrule}{ \noindent \parbox{\textwidth}{
        \hfill\rule[-3pt]{0.1mm}{3pt}\rule{0.5\textwidth}{0.1mm}} }

\def\bbbc{{\mathchoice {\setbox0=\hbox{$\displaystyle\rm C$}\hbox{\hbox
to0pt{\kern0.4\wd0\vrule height0.9\ht0\hss}\box0}}
{\setbox0=\hbox{$\textstyle\rm C$}\hbox{\hbox
to0pt{\kern0.4\wd0\vrule height0.9\ht0\hss}\box0}}
{\setbox0=\hbox{$\scriptstyle\rm C$}\hbox{\hbox
to0pt{\kern0.4\wd0\vrule height0.9\ht0\hss}\box0}}
{\setbox0=\hbox{$\scriptscriptstyle\rm C$}\hbox{\hbox
to0pt{\kern0.4\wd0\vrule height0.9\ht0\hss}\box0}}}}
\begin{document}

\title{Josephson scanning tunneling microscopy.}
\author{Jurij \v{S}makov$^{1,2}$\cite{smakov}, Ivar Martin$^2$ 
and Alexander V. Balatsky$^2$}
\address{$^1$Theoretical Physics, Department of Physics, Royal Institute of
Technology, SE-10044 Stockholm, Sweden\\
$^2$Theoretical Division, Los Alamos National Laboratory,
Los Alamos, NM 87545}

\date{Printed \today}

\maketitle

\begin{abstract}
We propose a set of scanning tunneling microscopy experiments in which
the surface of superconductor is scanned by a \emph{superconducting} tip.
Potential capabilities of such experimental setup are discussed. Most important
anticipated results of such an experiment include the position-resolved
measurement of the superconducting order parameter and the possibility to
determine the nature of the secondary component of the order parameter
at the surface. The theoretical description based on the tunneling
Hamiltonian formalism is presented. 
\end{abstract}


\begin{multicols}{2}


\section{Introduction}
Recent technological advances allowed Pan \emph{et al.} to conduct very
low temperature scanning tunneling microscopy (STM) experiments reaching
temperatures as low as 220 mK\cite{davis1} and making possible the direct
imaging of the surface of high-$T_c$ superconductors (SC) in a
superconducting state with high spatial and energy resolution.\cite{davis2}
The Pt/Ir tips used in these experiments were not superconducting, but
the same group has reported earlier,\cite{davis3} that they were able to
obtain the atomic resolution images of superconducting NbSe$_2$ with
the \emph{superconducting} atomically sharp Nb tip in the quasiparticle
tunneling regime. The idea of STM experiment in which the Josephson
effect would be observed between the tip and the surface was also
suggested.\cite{davis3}

If both the tip and the surface are superconducting and the temperature
is low enough to have a well-defined phase difference between the
superconductors, the tunneling contact may be
considered a Josephson junction with tunable parameters (hence the term
``Josephson STM'' or JSTM). The phase-dependent supercurrent will not be washed
out by thermal fluctuations if the temperature is sufficiently low,
\begin{equation}
\label{cond}
I_c\Phi_0 \gtrsim 2\pi k_BT,
\end{equation}
where $I_c$ is the critical Josephson current and 
$\Phi_0=h/2e$ is the unit magnetic flux. Assuming that the
Josephson current can be roughly estimated as 
$I=\sigma\Delta/e$,\cite{baratoff} 
where $\Delta$ is the magnitude of the superconducting order parameter (OP)
and $\sigma$ is the normal state conductance of the junction, 
Eq.\ (\ref{cond}) translates into the condition imposed on the normal state
resistance:
\begin{equation}
\label{res}
R \lesssim \frac{\Delta}{2 k_BT}R_0.
\end{equation}
Here $R_0=\hbar/e^2\approx 4\,\textrm{k}\Omega$ is the resistance quantum.
For conventional superconductors ($\Delta\sim 2\,\textrm{meV}$) and temperature
$T\sim 0.1\,\textrm{K}$ the junction resistance should not exceed 
$\sim 500\,\textrm{K}\Omega$ for the Josephson current to be observable.
For higher gap values, such as in high-$T_c$ materials,
we estimate $R\lesssim 7.5\,\textrm{M}\Omega$.
Since the temperature enters into criteria (\ref{res}) only via
ratio $\Delta/kT$, it might be easier to observe Josephson current
using high-$T_c$ material with a larger value of $\Delta$ as
a tip rather than to lower temperature by an order of magnitude.

Typical resistances of STM experiments are at least by an order of magnitude
larger, than the estimated upper bounds for Josephson effect. Thus, 
Josephson supercurrent is difficult to observe in a setup with an
atomically sharp superconducting tip. Therefore we propose the use of a
tip with a finite tunneling area. While it will be impossible to 
achieve the atomic-scale spatial resolution, to capture the
microscopic variations of superconducting OP only the resolution 
comparable to the coherence length of SC is needed. This characteristic length
is about 20 \AA\ for the high-$T_c$ materials and a few thousand \AA\ for
conventional superconductors. Also, the finite tunneling area will lead to
relatively big capacitance of the junction, minimizing the charging energy
and making the effects of Coulomb blockade negligible.

Since the supercurrent depends on the magnitudes of the superconducting OP
on both sides of the junction,\cite{baratoff} it should be
possible to measure the microscopic variations in OP depending on the position
of the tip. Such measurement can provide a direct insight into the
microscopic nature of the superconducting state. 

Another potential application of the JSTM is the probing of the 
\emph{symmetry} of superconducting OP. It is known that
a perturbation, such as an applied magnetic field or a surface, can induce a
secondary component of the superconducting OP. The type of the secondary
component can be determined from the modification of the Josephson current. 
In case when both the tip and the surface
are unconventional superconductors, like high-$T_c$ cuprates, the OP is
angle-dependent and the variations in the tunneling 
supercurrent may be caused by changing the mutual orientation, providing
the direct information about the symmetry of OP.

Finally, the JSTM could provide information about the nature of the gap in
high-$T_c$ superconductors. It is believed, that the normal state of the
underdoped high-$T_c$ cuprates is a pseudogap (PG) state with partially gapped
Fermi surface, from which superconductivity
emerges as the temperature is lowered through $T_c$.\cite{timusk}
Currently there is no general agreement on the origin of PG state. Some models
attribute PG to superconducting phase fluctuations above $T_c$,\cite{emery}
others - to a competing non-superconducting order parameter.\cite{chak}
Since JSTM is sensitive only to the superconducting OP, the measurements
of the critical Josephson current on differently doped samples can 
shed some light on the origin of PG.

The rest of the paper is organized as follows: in Section \ref{descr}
we describe the tunneling Hamiltonian formalism used in the calculations
and two practically important cases of JSTM, followed by proposed
experimental setup (Section \ref{setup}) and summary (Section \ref{summary}).

\section{Theoretical description}
\label{descr}
We base our theoretical description of JSTM on the tunneling Hamiltonian
formalism.\cite{baratoff,cohen}
In this formalism, time-dependent Josephson current through the tunneling
contact is expressed as\cite{mahan}
\begin{equation}
\label{current}
I_J(t)=2e\,\textrm{Im}[\exp(-2ieVt/\hbar)\Phi_{\textrm{ret}}(eV)],
\end{equation}
where $V$ is the applied voltage and $\Phi_{\textrm{ret}}(eV)$ is the
voltage-dependent retarded correlation function, which in the limit
of the zero temperature has the form
\begin{equation}
\label{rcf}
\Phi_{\textrm{ret}}(eV)=\frac{1}{2}\sum_{\mathbf{k}\mathbf{p}}
T_{\mathbf{k}\mathbf{p}}T_{-\mathbf{k},-\mathbf{p}}
\frac{\Delta_{\mathbf{k}}^{*}\Delta_{\mathbf{p}}}{E_{\mathbf{k}}E_{\mathbf{p}}}
\times
\end{equation}
$$
\times\left(\frac{1}{eV+E_{\mathbf{k}}+E_{\mathbf{p}}+i\delta}-
\frac{1}{eV-E_{\mathbf{k}}-E_{\mathbf{p}}+i\delta}
\right).
$$
Here indices $\mathbf{k}$ and $\mathbf{p}$ refer to the quasiparticle momenta
of the tip and surface, respectively, $T_{\mathbf{k}\mathbf{p}}$
is the tunneling matrix element, $\Delta_{\mathbf{k}}$ and
$\Delta_{\mathbf{p}}$ are the
superconducting order parameters, $E_{\mathbf{k}}$ and $E_{\mathbf{p}}$
are the excitation energies, and the infinitesimal quantity $i\delta$
ensures the convergence of the integral. 

Using Eqs.\ (\ref{current}) and (\ref{rcf}) we can calculate the Josephson
current
for a tunneling contact of two superconductors. Unique feature of the
Josephson effect is that below a certain  threshold value, the
dc supercurrent can flow without any voltage drop on the contact. 
This threshold value -- critical Josephson
current -- is easily measured, because when the value of the current reaches
the critical value, the finite voltage drop abruptly
appears at the contact. If the
analytical expression for the critical current is known, its measurements
can provide the information about the magnitudes of OPs, its symmetry and
possibly other quantities of interest.

\subsection{$s$-$d$ junction}
\label{s-d}
First we consider the case when the STM tip is a conventional $s$-wave
SC (superconducting metal, like Nb) with momentum-independent real order
parameter $\Delta_{\mathbf{k}}\equiv\Delta_1$
and the surface under study is the $a$-$b$ plane of a $d$-wave unconventional
SC (e.g., YBCO or BSCCO). It is generally agreed,
that perturbation of the unconventional $d$-wave
superconductors, like magnetic field or the distortion of crystal
lattice caused by presence of the surface, induces a secondary component
of the superconducting OP.\cite{balatsky,fogel,kleiner,kouz} 
Tunneling between pure $s$- and $d$-wave superconductors is expected
to be zero by symmetry arguments, at least if one does not take into account
the higher order processes.\cite{tanaka} Experimental
evidence exists,\cite{kleiner,kouz,greene} that the secondary component in 
high-$T_c$ cuprates possesses $s$-wave symmetry. In order to determine the
dependence of the critical Josephson current both on the magnitude and the 
phase of the secondary OP component, we have chosen the following form for
the superconducting OP on the surface:
\begin{equation}
\Delta_{\mathbf{p}}\equiv \Delta_2 =\Delta^{(d)}_2\cos 2\varphi+
e^{i\alpha}\Delta^{(s)}_2.
\end{equation} 
Here $\Delta^{(d)}_2$ and $\Delta^{(s)}_2$ are the magnitudes of the
primary and secondary
components (respectively) and $\varphi$ is the azimuthal angle corresponding to
momentum $\mathbf{p}$ in the coordinate system associated with the surface
plane. By varying angle $\alpha$ the secondary component can be assigned
an arbitrary
phase: In particular $\alpha=0$ corresponds to pure $d+s$ symmetry and
$\alpha=\pi/2$, to $d+is$.
The excitation energies $E_{\mathbf{k}}$ and $E_{\mathbf{p}}$ are given by
familiar relations
\begin{equation}
E_{\mathbf{k}}=\sqrt{\xi_{\mathbf{k}}^2+\Delta_1^2}
\end{equation}
and
\begin{equation}
E_{\mathbf{p}}=\sqrt{\xi_{\mathbf{p}}^2+|\Delta_2|^2},
\end{equation}
where $\xi_{\mathbf{k}}$ and $\xi_{\mathbf{p}}$ are the single-particle
energies, measured with respect to corresponding chemical potentials.
We assume here that the tunneling matrix element
is independent of momenta, i.e. 
\begin{equation}
T_{\mathbf{k}\mathbf{p}}T_{-\mathbf{k},-\mathbf{p}}=|T|^2e^{i\phi}.
\end{equation}
Here $|T|^2$ is a constant, depending only on the geometry of the tunneling
contact (it drops exponentially with increasing distance
between tip and the surface) and $\phi$ is the phase difference,
giving rise to Josephson supercurrent.

In order to calculate the sums in Eq. (\ref{rcf}), we convert them into
integrals
over the energies $\xi_{\mathbf{k}}$, $\xi_{\mathbf{p}}$ and the
azimuthal angle $\varphi$. In the case of zero applied voltage,
the integral with respect to energies can be
calculated using the result due to Ambegaokar and Baratoff\cite{baratoff} to
yield
\begin{equation}
\label{rcfv0}
\Phi_{\textrm{ret}}(0)=\frac{\sigma_0}{2\pi e^2}e^{i\phi}\int_0^{2\pi}
\frac{\Delta_1\Delta_2}{\Delta_1+|\Delta_2|}K\left(
\frac{|\Delta_1-|\Delta_2||}{\Delta_1+|\Delta_2|}\right)\,d\varphi
\end{equation}
Here $K(x)$ is the complete elliptic integral of the first kind and
$\sigma_0$ is the conductance of the tunneling contact in the normal
state\cite{mahan}
\begin{equation}
\sigma_0=4\pi e^2 N_1N_2|T|^2
\end{equation}
with $N_1$ and $N_2$ being the densities of states at the Fermi levels of
tip and surface respectively.
The critical Josephson current in this case is
\begin{equation}
I_c=I_c^0|\chi(\gamma,\kappa,\alpha)|,
\end{equation}
where $\gamma=\Delta^{(d)}_2/\Delta_1$ and $\kappa=\Delta^{(s)}_2/\Delta_1$
are dimensionless parameters, $I_c^0=\Delta_1\sigma_0/\pi e$ and
$\chi(\gamma,\kappa,\alpha)$ is the value of integral in Eq. (\ref{rcfv0}).
Assuming, that $\Delta_1$ is fixed, we have studied numerically the
dependence of the critical current on $\gamma$, $\kappa$ and $\alpha$ for
realistic parameter ranges.

\begin{figure}[htbp]
  \begin{center}
  \epsfxsize=3.0in
  \epsfbox{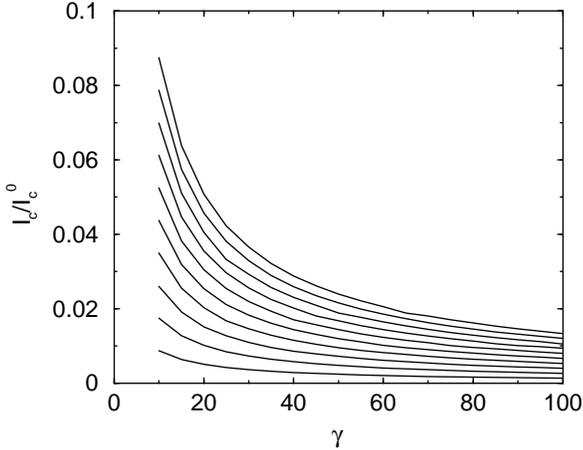}
\caption{The dependence of the critical Josephson current on $\gamma$
and $\kappa$ for an $s$-$d$ junction in the limit of zero applied
voltage and $\alpha=0$. Different curves correspond to different values 
of $\kappa$, starting from $\kappa=0.01$ (lowest curve) with the
increment of 0.01. The uppermost curve thus corresponds to $\kappa=0.1$.}
\label{s-d-alpha0}
\end{center}
\end{figure}

The dependence of the critical Josephson current for $\alpha=0$ (corresponds
to pure $d+s$ symmetry) on $\gamma$ and $\kappa$ is shown on Fig. 
\ref{s-d-alpha0}. 

\begin{figure}[htbp]
  \begin{center}
  \epsfxsize=3.0in
    \epsfbox{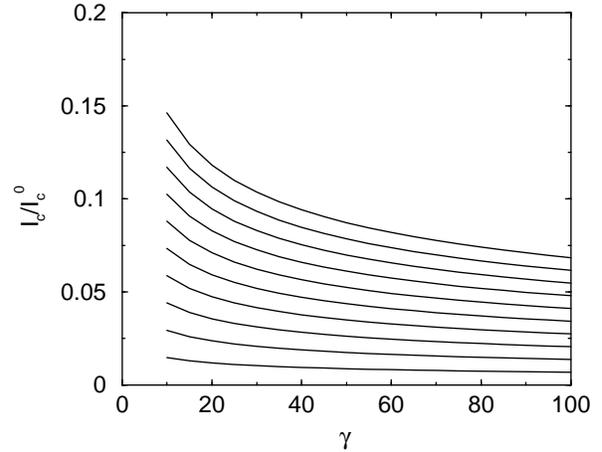}
\caption{The dependence of the critical Josephson current on $\gamma$
and $\kappa$ for an $s$-$d$ junction in the limit of zero applied
voltage and $\alpha=\pi/2$. Different curves correspond to different values 
of $\kappa$, starting from $\kappa=0.01$ (lowest curve) with the
increment of 0.01. The uppermost curve thus corresponds to $\kappa=0.1$.}
\label{s-d-alphapi2}
\end{center}
\end{figure}

\vspace{-1.5cm}
Another important case is $\alpha=\pi/2$, which corresponds to $d+is$
symmetry of superconducting OP. The results for this case are presented in 
Fig. \ref{s-d-alphapi2}.

\begin{figure}[htbp]
  \begin{center}
  \epsfxsize=3.0in
    \epsfbox{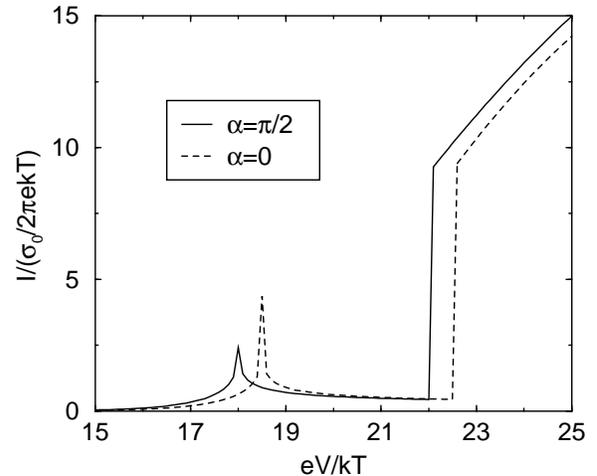}
\caption{Typical quasiparticle current-voltage characteristic (values
of parameters are given in the text) of a tunneling contact between
$d$-wave superconductor with the secondary OP component and $s$-wave
superconductor. Different values of $\alpha$ correspond to $d$+$s$
($\alpha=0$) and $d$+$is$ ($\alpha=\pi/2$) symmetries.}
\label{quasi}
\end{center}
\end{figure}

\vspace{-1.5cm}
Useful information can also be extracted from the observation of quasiparticle
current between superconductors at a finite temperature.
The current can be written as\cite{mahan,tinkham}
\end{multicols}
\lrule
\begin{equation}
I=\frac{\sigma_0}{2\pi e}\int_0^{2\pi}\,d\varphi\int_{-\infty}^{+\infty}
\frac{|E|}{\sqrt{E^2-\Delta_1^2}}\frac{|E+eV|}{\sqrt{(E+eV)^2-|\Delta_2|^2}}
[f(E)-f(E+eV)]\,dE,
\end{equation}
\rrule
\begin{multicols}{2}\noindent
where $f(E)$ is a Fermi-Dirac distribution function and the regions 
$|E|<\Delta_1$ and $|E+eV|<|\Delta_2|$ are excluded from the integration.
We have numerically
calculated current-voltage characteristic for a realistic
set of parameters: $\Delta_1/kT=2$, 
$\Delta_2^{(d)}/\Delta_1=10$ and $\Delta_2^{(s)}/\Delta_2^{(d)}=0.025$. It is
known,\cite{tinkham} that when the tunneling contact of two $s$-wave
superconductors with different constant real order parameters $\Delta_1$
and $\Delta_2$ is considered, the quasiparticle current-voltage
characteristic possesses two significant features: a peak at the voltage
$|\Delta_1-\Delta_2|$ and an abrupt increase in current at voltage
$\Delta_1+\Delta_2$, corresponding to transition to Ohmic regime.
We find that the position of the peak and the Ohmic
transition depends on the symmetry of the superconducting OP. In particular
(as seen in Fig. \ref{quasi}) the current is sharply peaked at 
the voltage $\Delta_2^{(d)}-\Delta_1$ for $\alpha=\pi/2$ ($d$+$is$ symmetry),
while for $\alpha=0$ the peak is at the voltage $\Delta_2^{(d)}+\Delta_2^{(s)}-
\Delta_1$. The transition to Ohmic regime is also shifted to higher voltages
by the value of $\Delta_2^{(s)}$. Thus, such a measurement can complement
the measurements of Josephson critical current to provide additional
information about the nature and magnitude of superconducting OP on the
surface.

\vspace{-1.5cm}
\subsection{$d$-$d$ junction}
\label{d-d}
The $d$-wave nature of a pair requires finite size tunneling area.
If the tunneling occurs from only one site on the tip it would imply 
effective momentum averaging of pair wavefunction and would yield zero
Josephson current for $d$-wave pairs. Thus it is desirable to have a finite
region of the tip where tunneling can occur. In this case, one looses
spatial resolution, compared to conventional tips. However the gain is
the phase coherent current between the tip and the surface. While the
idea of using the high-$T_c$ cuprates as a material of STM tip has not
yet been reported in the literature, the fabrication of relatively small
(of the order 100 \AA) flat high-$T_c$ probes, suitable for use as
tips, is feasible. 

Theory based on tunneling Hamiltonian predicts, that if the tunneling
matrix element is momentum-independent, $c$-axis Josephson current at
zero temperature is identically
zero for superconductors with pure $d$-wave symmetry of OP. It is
possible to argue that, as
discussed in previous section, the presence of the surface will induce the
secondary OP component, which can lead to a nonzero current. However, since the
magnitude of secondary component is usually much smaller than the magnitude
of primary one, this induction will produce a second order effect and we
assume that it can be neglected. 
\begin{figure}[htbp]
  \begin{center}
  \epsfxsize=3.0in
    \epsfbox{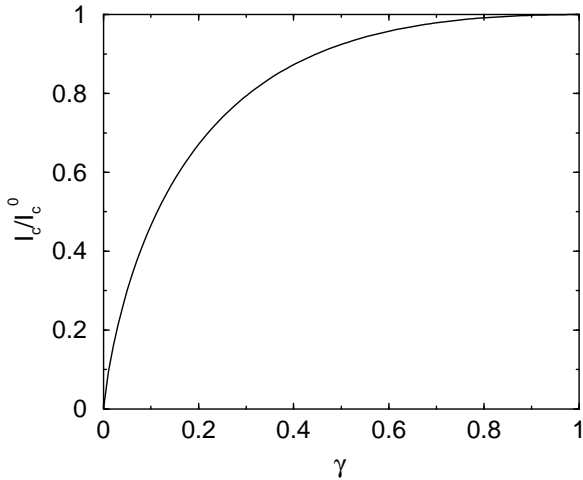}
\caption{The dependence of the critical Josephson current on the ratio
of OP magnitudes for a $d$-$d$ junction in the limit of zero applied
voltage and zero rotation.}
\label{gaps}
\end{center}
\end{figure}

Here we discuss a different scenario, supported by recent theoretical and
experimental work of Latyshev \emph{et al.}\cite{latysh} In their work, the
\emph{intrinsic} $c$-axis quasiparticle transport between
layers of BSCCO was studied. It was found that interlayer transport is
predominantly momentum-conserving, i.e. the in-plane components of
quasiparticle momentum are conserved in the tunneling process. It seems
natural to make an assumption, that in-plane momentum is also conserved
in case of Josephson supercurrent. As we show below, adoption of this
hypothesis leads to non-zero current even in the case of pure $d$-wave
symmetry.

We consider an $a$-$b$ plane tunneling contact of two $d$-wave
superconductors. We allow the magnitudes of the superconducting OP
to be different. 
Because of the angular dependence of the superconducting OP, the Josephson
current will depend on the mutual orientation of two superconductors and
we introduce angle $\theta$ (angle between $a$-axes of superconductors)
to describe this orientation. The problem is essentially two-dimensional
and the order parameters of the tip $\Delta_1$ and surface $\Delta_2$
may be written as
\begin{equation}
\Delta_1=\Delta^{(0)}_1\cos 2\varphi
\end{equation}
and
\begin{equation}
\Delta_2=\Delta^{(0)}_2\cos 2(\varphi+\theta).
\end{equation}
Tunneling matrix element has the form
\begin{equation}
T_{\mathbf{k}\mathbf{p}}T_{-\mathbf{k},-\mathbf{p}}=|T|^2e^{i\phi}
\delta(\mathbf{k}-\mathbf{p}),
\end{equation}
where $\mathbf{k}$ and $\mathbf{p}$ are two-dimensional vectors. Again,
we change the sums in Eq. (\ref{rcf}) into integrals, which can be calculated
numerically. In the case of zero applied voltage and zero rotation
angle $\theta$ it is possible to calculate the integral analytically and
find that the critical Josephson current in this limit is
\begin{equation} 
I_c=I_c^0\frac{\gamma\ln \gamma^2}{\gamma^2-1},
\end{equation}
where $\gamma=\Delta^{(0)}_1/\Delta^{(0)}_2$, assuming 
$\Delta^{(0)}_1<\Delta^{(0)}_2$, and
$I_c^0$ is constant current independent of $\gamma$. This dependence
is presented in Fig. \ref{gaps}. $\gamma=0$ corresponds to absence of the
superconductivity in the tip, which leads to dissapearance of the Josephson
current.

\begin{figure}[htbp]
  \begin{center}
  \epsfxsize=3.0in
    \epsfbox{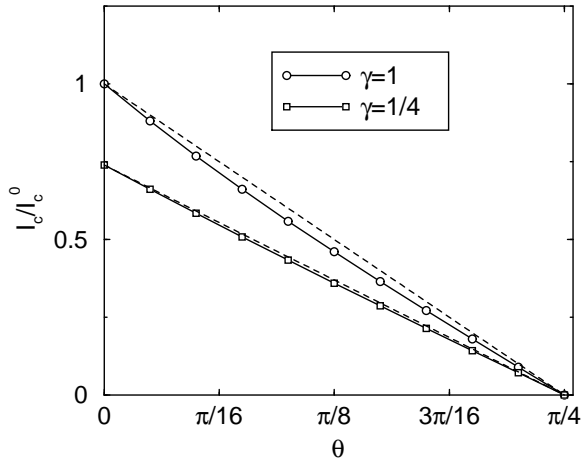}

\caption{The dependence of the critical Josephson current on the rotation
angle $\theta$ for a $d$-$d$ junction in the limit of zero applied
voltage at different values of $\gamma$ (solid lines). Dashed lines
show the linear approximation of Eq. (\ref{linear}).}
\label{rot}
\end{center}
\end{figure}

We have also studied the dependence of the critical current on the rotation
angle $\theta$ at zero applied voltage (Fig. \ref{rot}).
As expected, it has a maximum at $\theta=0$, monotonically
decreases with increasing $\theta$ and becomes zero at $\theta=\pi/4$. 
For the angles bigger than $\pi/4$, the critical current can be obtained
as 
\begin{equation}
I_c\left(\frac{\pi}{4}+\theta\right)=-I_c\left(\frac{\pi}{4}-\theta\right),
\end{equation}
i.e. for $\theta>\pi/4$ the current changes direction, but the curve
remains symmetric with respect to the point $\theta=\pi/4$. It is notable,
that since the curve is symmetric with respect to $\theta=0$,
there is a kink in the angle dependence at $\theta=0$.

It can be seen from Fig. \ref{rot}, that the decrease is almost
linear with $\theta$
and linear approximation becomes better as $\gamma$ is reduced. For practical
purposes an approximate formula 
\begin{equation}
\label{linear}
I_c=I_c^0\frac{\gamma\ln\gamma^2}{\gamma^2-1}\left(1-\frac{4}{\pi}\theta
\right)
\end{equation}  
may be used.

It is worth mentioning that when the angle $\theta$ is non-zero, the
denominator in Eq. (\ref{rcf}) has a positive lower bound. This means that for
voltages much smaller than this lower bound, the integrand may be expanded
in powers of voltage. It turns out that all first order terms in voltage
exactly cancel out and dependence on voltage enters
only as higher order corrections (proportional to the square of the voltage),
so the zero-voltage behavior is not altered significantly. The range of
applicability of this approximation depends on actual values of $\gamma$ and
$\theta$.

\section{Experimental setup}
\label{setup}
\begin{figure}[htbp]
  \begin{center}
  \epsfxsize=3.0in
    \epsfbox{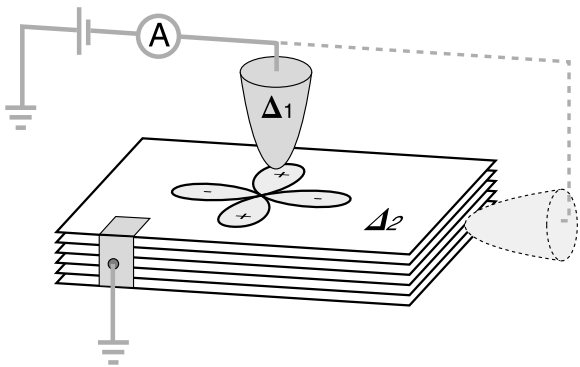}
\caption{Proposed experimental setup for $s$-$d$ tunneling.}
\label{s-dexp}
\end{center}
\end{figure}

\begin{figure}[htbp]
  \begin{center}
  \epsfxsize=3.0in
    \epsfbox{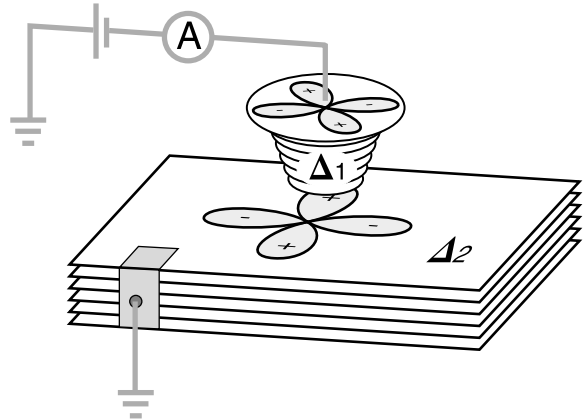}
\caption{Proposed experimental setup for $d$-$d$ tunneling.}
\label{d-dexp}
\end{center}
\end{figure}

The sketches of experimental setup for $s$-$d$ and $d$-$d$ tunneling 
are shown in Fig. \ref{s-dexp} and
Fig. \ref{d-dexp} respectively. In both cases, superconducting tip and
surface form an electric circuit together with current (voltage) source
and measuring equipment.

In $s$-$d$ case, measurements can be performed in two stages. First, by
driving a large current through the tunneling contact, the superconductivity
is suppressed and the normal state conductance of the tunneling junction
at the specific position on the surface is measured. Next, the measurement
of the critical Josephson current (by starting with zero current and increasing
it until a voltage drop appears on the contact) is performed.
Using this information, a conclusion can be made about the magnitudes of the
primary and/or secondary OP components on the surface. By repeating the
measurement, a surface map of variations in order parameter may be obtained.
Both the surface, formed by $a$-$b$ plane and the one, parallel to $c$-axis
can be studied.
In $d$-$d$ case, the apparatus must be designed in such a way that the tip
can be rotated around vertical axis.
Measuring the Josephson current as a function of rotation angle, the
information about the symmetry of OP can be extracted. If Josephson current
has nodes, separated by $\Delta\theta=\pi/2$, a conclusion can be made
that the non-zero Josephson current is due to coherent
tunneling and superconductors possess pure $d$-wave symmetry. 
If, on the other hand, current never becomes zero or becomes zero
with different periodicity, then the secondary component of the OP must be a
primary reason for the existence of the current.

Finally, the experiments aimed at establishing the nature of the pseudogap
should be conducted with differently doped tip and surface. For example,
measuring critical Josephson current in a setup with underdoped tip and
overdoped surface and comparing it with measurements on identically underdoped
(or overdoped) contacts will provide information about the magnitude
of the pseudogap and whether or not it is of superconducting origin. Detailed
discussion of such an experiment will be given elsewhere. 

\section{Summary}
\label{summary}
In summary, we have a proposed a series of STM experiments of a
superconducting surface with a superconducting tip. Treating the tunneling
contact as a Josephson junction with tunable parameters, we have developed
a theoretical description of these experiments, calculating the dependence
of critical Josephson current on various relevant parameters. Since Josephson
current is sensitive to the magnitudes of superconducting order parameters
in the tip and in the surface, this new technique allows to measure the
microscopic spatial variations of the OP on the surface and determine the
magnitude and phase of the secondary OP component, induced by the surface.
It is also possible to probe the symmetry of superconducting OP directly,
exploiting the dependence of the Josephson critical current on the mutual
orientation of $d$-wave superconductors. Finally, since the Josephson current
is sensitive only to the superconducting gap, the new technique can provide
information about the nature of the pseudogap, in particular, confirm or
falsify its superconducting origin.

\section*{Acknowledgments}

We are grateful to J. C. Davis, M. J. Graf and A. Rosengren
for many valuable comments.
One of us (J. \v{S}.) would like to thank Los Alamos National Laboratory for
hospitality and acknowledge partial financial support from Swedish Natural
Science Research Council.

\end{multicols}

\begin{thebibliography}{99}
\bibitem[*]{smakov}
Corresponding author.\\ Electronic address: {\tt jurijus@theophys.kth.se}.
\bibitem{davis1}
S. H. Pan, E. W. Hudson and J. C. Davis, Rev. Sci. Instrum.
{\bf 70} (2), 1459 (1999).
\bibitem{davis2}
S. H. Pan \emph{et al.}, Nature {\bf 403}, 746 (2000); 
S. H. Pan \emph{et al.}, Phys. Rev. Lett. {\bf 85}, 1536 (2000).
\bibitem{davis3}
S. H. Pan, E.W. Hudson and J. C. Davis, Appl. Phys. Lett.
{\bf 73}, 2992 (1998).
\bibitem{baratoff}
V. Ambegaokar and A. Baratoff, Phys. Rev. Lett. {\bf 10}, 486 (1963);
\emph{ibid.} {\bf 11}, 104 (E) (1963).
\bibitem{timusk}
T. Timusk and B. Statt, Rep. Prog. Phys. {\bf 62}, 61 (1999).
\bibitem{emery}
V. J. Emery and S. A. Kivelson, Nature {\bf 374}, 434 (1995).
\bibitem{chak}
P. A. Lee and X. G. Wen, Phys. Rev. Lett. {\bf 76}, 503 (1996);
I. Martin, G. Ortiz, A. V. Balatsky and A. R. Bishop, cond-mat/0003316;
S. Chakravarty, R. B. Laughlin, D. K. Morr and C. Nayak, cond-mat/0005443.
\bibitem{cohen} 
M. H. Cohen, L. M. Falicov and J. C. Phillips, Phys. Rev. Lett. {\bf 8},
316 (1962).
\bibitem{mahan}
G. D. Mahan, \emph{Many-Particle Physics} (Plenum Press, New York, 1990).
\bibitem{balatsky} K. Krishana, et al., Science {\bf 277}, 83 (1997); 
R. B. Laughlin, Phys. Rev. Lett. {\bf 80}, 5188 (1998);
A. V. Balatsky, P. Kumar and J. R. Schrieffer, Phys. Rev. Lett. {\bf 84},
4445 (2000) and references therein.
\bibitem{fogel}
M. Fogelstr\"om, D. Rainer and J. A. Sauls, Phys. Rev. Lett. {\bf 79},
281 (1997)
\bibitem{kleiner}
R. Kleiner \emph{et al.}, Phys. Rev. Lett. {\bf 76}, 2161 (1996).
\bibitem{kouz}
K. A. Kouznetsov \emph{et al.}, Phys. Rev. Lett. {\bf 79}, 3050 (1997).
\bibitem{tanaka}
Y. Tanaka, Phys. Rev. Lett., {\bf 72}, 3871 (1994).
\bibitem{greene}
M. Covington \emph{et al.}, Phys. Rev. Lett. {\bf 79}, 277 (1997). 
\bibitem{tinkham}
M. Tinkham, \emph{Introduction to Superconductivity} (McGraw-Hill, 
New York, 1996).
\bibitem{latysh}
Yu. I. Latyshev \emph{et al.}, Phys. Rev. Lett. {\bf 82}, 5345 (1999).
\end{thebibliography}
\end{document}